\documentstyle[12pt,epsf]{article} \hoffset -0.5in \textwidth 6.00in
\textheight
8.5in  \parskip 7pt \openup1.5\jot \parindent=0.5in
\newskip\humongous \humongous=0pt plus 1000pt minus 1000pt
\def\caja{\mathsurround=0pt}
\def\eqalign#1{\,\vcenter{\openup1\jot \caja
        \ialign{\strut \hfil$\displaystyle{##}$&$
        \displaystyle{{}##}$\hfil\crcr#1\crcr}}\,}
\newif\ifdtup

\def\eqright #1\cr{\noalign{\hfill$\displaystyle{{}#1}$}}
\def\eqleft #1\cr{\noalign{\noindent$\displaystyle{{}#1}$\hfill}}

\def\oldreffmt#1{\rlap{[#1]} \hbox to 2\parindent{}}

\def\figfmt#1{\rlap{Figure {#1}} \hbox to 1in{}}

\newcounter{holdequation}

\def\auto{\eqno(\refstepcounter{equation}\theequation)}
\def\begineq #1\endeq{$$ \refstepcounter{equation}\eqalign{#1}\eqno
	(\theequation) $$}
\def\contlimit{\,{\hbox{$\longrightarrow$}\kern-1.8em\lower1ex
\hbox{${\scriptstyle (a\rightarrow0)}$}}\,}
\def\centeron#1#2{{\setbox0=\hbox{#1}\setbox1=\hbox{#2}\ifdim
\wd1>\wd0\kern.5\wd1\kern-.5\wd0\fi
\copy0\kern-.5\wd0\kern-.5\wd1\copy1\ifdim\wd0>\wd1
\kern.5\wd0\kern-.5\wd1\fi}}
\def\centerover#1#2{\centeron{#1}{\setbox0=\hbox{#1}\setbox
1=\hbox{#2}\raise\ht0\hbox{\raise\dp1\hbox{\copy1}}}}
\def\centerunder#1#2{\centeron{#1}{\setbox0=\hbox{#1}\setbox
1=\hbox{#2}\lower\dp0\hbox{\lower\ht1\hbox{\copy1}}}}
\def\lsim{\;\centeron{\raise.35ex\hbox{$<$}}{\lower.65ex\hbox
{$\sim$}}\;}
\def\gsim{\;\centeron{\raise.35ex\hbox{$>$}}{\lower.65ex\hbox
{$\sim$}}\;}

\def\super#1{\ifmmode \hbox{\textsuper{#1}}\else\textsuper{#1}\fi}
\def\textsuper#1{\newcount\holdspacefactor\holdspacefactor=\spacefactor
$^{#1}$\spacefactor=\holdspacefactor}

\def\getcite#1,{\advance\citenumber by1
\ifnum\citenumber=1
\ref{#1}\let\next=\getcite\else\ifx#1@\let\next=\relax
\else ,\ref{#1}\let\next=\getcite\fi\fi\next}

\def\upon #1/#2 {{\textstyle{#1\over #2}}}
\relax
\renewcommand{\thefootnote}{\fnsymbol{footnote}}

\def\subhead#1{\bigskip\vbox{\noindent\bf #1}\nobreak\par}

\def\til#1{\centeron{\hbox{$#1$}}{\lower 2ex\hbox{$\char'176$}}}
\def\tild#1{\centeron{\hbox{$\,#1$}}{\lower 2.5ex\hbox{$\char'176$}}}
\def\sumtil{\centeron{\hbox{$\displaystyle\sum$}}{\lower
-1.5ex\hbox{$\widetilde{\phantom{xx}}$}}}

\def\pom{{\rm P\kern -0.53em\llap I\,}}
\def\spom{{\rm P\kern -0.36em\llap \small I\,}}
\def\sspom{{\rm P\kern -0.33em\llap \footnotesize I\,}}

\begin{document}
\begin{titlepage}
\rightline{\vbox{\halign{&#\hfil\cr
&ANL-HEP-CP-94-79\cr
&\today\cr}}}
\vspace{0.25in}
\begin{center}

{\large\bf
SCALE-INVARIANT LIPATOV KERNELS FROM t-CHANNEL UNITARITY}

\medskip

Claudio Corian\`{o} and Alan R. White
\footnote{Work supported by the U.S. Department of
Energy, Division of High Energy Physics, Contract\newline W-31-109-ENG-38}
\\ \smallskip
High Energy Physics Division, Argonne National Laboratory, Argonne, IL
60439.\\ \end{center}

\begin{abstract}
The Lipatov equation can regarded as a reggeon Bethe-Salpeter equation in
which higher-order reggeon interactions give higher-order kernels. Infra-red
singular contributions in a general kernel are produced by t-channel
nonsense states and the allowed kinematic forms are determined by unitarity.
Ward identity and infra-red finiteness gauge invariance constraints then
determine the corresponding scale-invariant part of a general higher-order
kernel.

\end{abstract}

\vspace{2in}

\noindent Presented at the Summer Institute on QCD, Gran Sasso
(Aug 29 - Sept 11), the XXIV International
Symposium on Multiparticle Dynamics, Vietri sul Mare (Sept 12-19), and the
2nd Workshop on Small-x and Diffractive Physics at the Tevatron, Fermilab
(Sept 22-24).

\renewcommand{\thefootnote}{\arabic{footnote}} \end{titlepage}

\subhead{1. INTRODUCTION}

The small-x behavior of parton disributions has been the cause of much
excitement. In particular, the ``BFKL Pomeron''\cite{lip}, i.e.
$$
\eqalign { F_2(x,q^2) ~\sim ~ x^{1-\alpha_0} ~\sim ~x^{-{1 \over 2}}}
\auto
$$
may have been seen at HERA. This behavior is obtained by solving the Lipatov
equation\cite{lip}, written as an evolution equation for parton
distributions, i.e.
$$
\eqalign{ {\partial \over \partial (ln {1 / x})}F(x,k^2) ~=~\tilde{F}(x,k^2)~+~
{1 \over (2\pi)^3}\int {d^2k' \over (k')^4} ~K(k,k') F(x,(k')^2) }
\auto\label{eve}
$$
where $K(k,q)~=~K^{(2)}_{2,2}(k,-k,q,-q)$ and $K^{(2)}_{2,2}$ is the full
$O(g^2)$ Lipatov kernel defined below. To obtain non-leading corrections to
the $O(g^2)$ kernel (which determine the crucial corrections to $\alpha_0$)
very complicated non-leading log Regge limit calculations are required.
Can general Regge theory help?

Our answer is, of course, yes and our purpose in this talk is to show that
the ``scale-invariant'' part of the $O(g^{2N})$ kernel, in which the gauge
coupling does not run and there is no transverse momentum scale, is actually
determined by the combination of Regge theory with gauge invariance. That is
by the combination of multiparticle $t$-channel unitarity, continued in the
j-plane\cite{arw1}, with Ward identity and infra-red finiteness constraints.
We will outline a different, but closely related, method to that given in
\cite{ker} for the explicit construction of higher-order kernels. Our
methods are t-channel based and distint from the application of s-channel
unitarity used by Bartels\cite{jb} to obtain higher-order results, although our
results do overlap.

\subhead{2. REGGEON FORMALISM}

To introduce reggeon language we rewrite the Lipatov equation as a
``reggeon Bethe-Salpeter equation". In effect we work backwards
historically. We first extend (\ref{eve}) to the non-forward direction, then
transform to $\omega$ - space (where $\omega$ is conjugate to $ln~{1 \over
x}$), giving
$$
\eqalign{ \omega F(\omega,k,q-k) ~=~\tilde{F}~+~
{1 \over (2\pi)^3}\int {d^2k' \over (k')^2(k'-q)^2}~K(k,k',q)
F(\omega,k',q-k') }
\auto\label{ome}
$$
where $K(k,k',q)~=~K^{(2)}_{2,2}(k,q-k,k',q-k')$ now contains three kinematic
forms i.e.
$$
\eqalign{& {2 \over 3g^2}K^{(2)}_{2,2}(k_1,k_2,k_3,k_4)~~~~~\equiv
{}~~~~~~~~~~K_1~+~K_2~+K_3 ~~~~~~\equiv ~~~\sum_{\scriptscriptstyle 1<->2}\cr
&\Biggl({1 \over 2} (2\pi)^3k_1^2J_1(k_1^2)k_2^2\Bigl(k_3^2\delta^2(k_2-k_4)
+k_4^2\delta^2(k_2-k_3)\Bigr)
-{k_1^2k_4^2~+~k_2^2k_3^2 \over (k_1-k_3)^2}
-(k_1+k_2)^2\Biggr)}
\auto\label{2,2}
$$
and
$$
\eqalign{J_1(q^2)~=~{1 \over (2\pi)^3}\int {d^2k' \over (k')^2(k'-q)^2}}
\auto
$$

Moving the $K_1$ term to the left side of (\ref{ome}) and defining
$G~=~\Gamma_2F$ gives
$$
\eqalign{ G(\omega,k,q-k) ~=~\tilde{G}+
{1 \over (2\pi)^3}\int {d^2k' \over (k')^2(k'-q)^2}~\Gamma_2(\omega,k',q-k')
\tilde{K}(k,k',q) G(\omega,k',q-k') }
\auto\label{bet}
$$
where $\Gamma_2(\omega,k_1,k_2)~=~[\omega
-g^2k_1^2J_1(k_1^2)-g^2J_1(k_2^2)]^{-1}$, is a
2-reggeon propagator and
$$
\eqalign{\tilde{K}(k,k',q)~=~ K_2~+K_3~ =~\sum_{\scriptscriptstyle 1<->2}
\Biggl({k_1^2k_4^2~+~k_2^2k_3^2 \over (k_1-k_3)^2}
-(k_1+k_2)^2\Biggr)}
\auto\label{int}
$$
is a 2-2 reggeon interaction. That this interaction is singular is what
makes it, at first sight, hard to understand from a Regge theory point of
view.

If we write ${1 \over k^2} \sim {1 \over sin \pi \alpha'k^2}$ and arrange
that $\alpha'$ scales out of the theory, then, apart from the singular
interaction, (\ref{bet}) is a simple two-reggeon Bethe-Salpeter
equation for odd-signature reggeons\cite{jb,reg}. In fact we can also include
the $K_1$ term in the
interaction and still write the reggeon equation (\ref{bet}) if we take
$\Gamma_2(\omega,k_1,k_2)~=~[\omega -\alpha'k_1^2-\alpha;k_2^2]^{-1}$, and
ultimately take $\alpha' \to 0$. In this way we can also interpret
(\ref{ome}) directly as a reggeon equation in which the interaction (the full
$K^{(2)}_{2,2}$ ) then has two vital properties

\begin{itemize}

\item It is infra-red finite as an integral kernel i.e.
$$
\eqalign{ \int {d^2k_1 \over k_1^2} {d^2k_2 \over k_2^2} \delta^2(q-k_1-k_2)
K^{(2)}_{2,2}(k_1,k_2,k_3,k_4) ~~~~~~is ~~finite}
\auto\label{fin}
$$

\item It contains singularities (poles) but satisfies the Ward
identity constraint
$$
\eqalign{ K^{(2)}_{2,2}(k_1,k_2,k_3,k_4) ~~\to ~~0~~, k_i~\to~ 0~, i=~1,..,4}
\auto\label{war}
$$

\end{itemize}

\noindent These two properties determine the relative magnitude of the three
kinematic forms $K_1, K_2,$ and $K_3$. If we can derive their existence from
a general Regge theory argument, as we do below, then the above properties
determine the kernel uniquely.

It will be convenient to introduce a diagrammatic notation for transverse
momentum integrals. A vertex with $n$ incoming and $m$ outgoing lines
represents
$$
\eqalign {(2\pi)^3\delta^2(\sum k_i~  - \sum k_i')(\sum k_i~)}
\auto
$$
and an $n$ line intermediate state represents
$$
\eqalign{ (1/2\pi)^{3n}\int d^2k_1...d^2k_n~ /~k_1^2...k_n^2 .}
\auto
$$
We also define all kernels to include a factor
$(2\pi)^3\delta^2(\sum k_i-\sum k_i')$, (they are then dimensionless and
formally scale-invariant). We can then represent (\ref{2,2}) as in Fig.~1,
where the sum is over all distinct momentum permutations for all
of the diagrams. We shall use these diagrams extensively in the following.

The higher-order corrections we want are evidently higher-order reggeon
interactions. We will try to determine them by generalising the above
discussion. As we have described in \cite{ker}, generalisations of
(\ref{fin}) and (\ref{war}) are properties that should be satisfied by any
(color zero) reggeon amplitude as a manifestation of gauge invariance. Our
purpose now is to argue that the singular kinematic forms that can be
present in a general reggeon interaction are actually determined by
multiparticle t-channel unitarity continued in the $\omega$ or ``$j$'' -
plane ($\omega = j - 1$ ). We shall then argue that (\ref{fin}) and
(\ref{war}) are sufficient to determine the scale-invariant part of a
general interaction.

\subhead{3. MULTIPARTICLE UNITARITY IN THE j - PLANE}

It is well-known that multiparticle $t$-channel unitarity can be used to derive
Regge cut discontinuities\cite{arw1}. We can very briefly illustrate this as
follows. Consider the four-particle intermediate state and insert Regge
poles in the production amplitude as in Fig.~2. If we initially ignore the
subtleties of signature we can write the relevant part of the j-plane
continuation of this equation as a (double) helicity integral of the form
$$
\eqalign{a_j - a_j^* = \int d\rho(t,t_1,t_2)
 \int {dn_1dn_2 \over sin\pi n_1 sin\pi n_2
sin\pi(j- n_1-n_2)} {A^+A^- \over (n_1-\alpha_1)  (n_2-\alpha_2)} }
\auto\label{jpl}
$$
where $a_j$ is the elastic partial-wave amplitude, $A^+$ and $A^-$ are
production amplitude Regge pole residues and $\alpha_i \equiv \alpha(t_i), i
= 1,2$. When the Regge poles at $n_1=\alpha_1$ and $n_2=\alpha_2$ combine
with the ``nonsense'' pole at $j=n_1-n_2-1$ to pinch the $n_1$ and $n_2$
integration contours in (\ref{jpl}) we obtain
$$
\eqalign{\int { d\rho \over (j - \alpha_1 -  \alpha_2)}
 {A^+A^- \over (sin\pi \alpha_1)  (sin\pi \alpha_2)} }
\auto\label{rct}
$$
which produces a Regge cut (at $j=2\alpha({t \over 4}) -1)$, {\bf provided}
there is no ``nonsense zero'' in $A$. The requirement of no zero actually
leads to the Regge cut appearing only in the even signature partial-wave
amplitude.

The gluon reggeon (with $\alpha' \neq 0$) is present in the odd
signature amplitude. In this amplitude the nonsense pole and the Regge poles in
(\ref{jpl}) combine to give
$$
\eqalign {sin\pi j~\int d\rho  {A^+A^- \over (sin\pi \alpha_1)
(sin\pi \alpha_2)} }
\auto\label{odd}
$$
where the $sin\pi j$ factor can be understood as originating from the
nonsense zero. If we now insert $ a_j \sim {1 \over (j-\alpha)}$ for $ j-1
\sim q^2 \to 0$ then (\ref{odd}) gives
$$
\eqalign { \alpha(q^2) - 1 ~ \sim~ q^2 \int { d\rho \over sin\pi \alpha_1
sin\pi \alpha_2} ~\sim~ q^2 \int {d^2k \over k^2(q-k)^2} }
\auto\label{rez}
$$
where now the $q^2$ factor providing the ``reggeization'' arises from the
nonsense zero. We see that reggeization is explicitly due to the
contribution of reggeon nonsense states.

If we move on to the six-particle intermediate state in the unitarity
equation we find that the three reggeon nonsense states give singular terms
in the 2-2 reggeon interaction i.e. the 2-2 Lipatov kernel. We will not give
the details here (they will be given in a future publication). Instead we
represent the construction as in Fig.~3. Clearly, if we add a constant 2-2
reggeon interaction, we generate the full set of transverse momentum
diagrams for the Lipatov kernel. If we impose the properties (\ref{fin}) and
(\ref{war}) above we then {\bf determine the full kernel from three-reggeon
nonsense states without calculating a single Feynman diagram.}

In leading-order higher multiparticle nonsense states give {\bf either}
multi-reggeon cuts, {\bf or} singularities of reggeon interactions i.e. {\bf
singular terms in higher-order kernels}. Note, however, that we can also apply
the odd-signature analysis of (\ref{odd}) and (\ref {rez}) to the even
signature channel contribution given by the non-leading behaviour of the
partial-waves at the nonsense pole. This implies that even signature states
will give nonsense-state contributions to reggeon interactions as a
non-leading effect.

\subhead{ 4. THE $O(g^4)$ KERNELS}

The main contribution to the $O(g^4)$ 2-2 kernel is from the four-particle
nonsense state, which we analyse via eight-particle unitarity. We consider all
couplings of two reggeons to the nonsense states involved and combine them
in all possible ways, as in Fig.~4. This gives the set of transverse
momentum diagrams that can be generated. Putting nonsense zeroes for the
vertices and imposing generalisations of (\ref{fin}) and (\ref{war}) then
gives $K^{(4)}_n$ uniquely. The diagrammatic representation of is given in
Fig.~5. The full $O(g^4)$ kernel also contains a contribution from
$(K^{(2)}_{2,2})^2$, where this kernel is defined as illustrated in Fig.~6.
This contribution was not considered in \cite{ker}. It is important, in
particular, because diagrams appear which are of the same form as the first
diagrams (i.e. the disconnected bubbles) appearing in the four-particle
nonsense state contribution above. These diagrams can not be associated with
the reggeization of either of the interacting reggeons. This implies they
have no Regge theory interpretation and must cancel. As a result the full
$O(g^4)$ 2-2 kernel is determined to be
$$
\eqalign{K^{(4)}_{2,2} ~=~(K^{(4)}_n~-~K^{(2)}_{2,2})^2}
\auto
$$
Using the above rules for diagrams gives the various contributions in the
form given in \cite{ker}.

We can also consider (N-M) reggeon interactions. At $O(g^4)$ the (2-4)
interaction appears and the relevant nonsense states are shown in Fig.~7.
Once again the gauge invariance constraints analagous to (\ref{fin}) and
(\ref{war}) uniquely determine the coefficients of the distinct transverse
momentum diagrams. The result for $K^{(4)}_{2,4}$ is the (completely
symmetrized) sum shown in Fig.~8. A complete expression for $K^{(4)}_{2,4}$
can be found in \cite{ker}. It is closely related to the kernel that appears
in deep-inelastic high-mass diffraction\cite{bar}.

\subhead{5. CONSTRUCTION RULES FOR GENERAL HIGHER-ORDER KERNELS}

{}From the above discussion it is clear that we can similarly construct a
general high-order arbitrary (N-M) kernel. The leading contribution to
$K^{(2N)}_{2,2}$ is from the (N+2) particle state. We draw all possible
point couplings and  combine them to form a full set of transverse momentum
diagrams as in Fig.~9.We also have to add all posible products of
lower-order kernels i.e.
$$
\eqalign{ (K^{(2)})^N~+~K^{(2)}~K^{(4)}(K^{(2)})^{N-3}~+~...}
\auto
$$
The cancellation of all disconnected diagrams that can not be interpreted in
terms of reggeization will produce a large number of constraints and we
anticipate that the collection of of Ward identity and infra-red finiteness
constraints associated with a wide range of distinct kinematic forms will
determine a unique scale-invariant kernel.

As a further example the contribution of the five-particle nonsense states
to the $O(g^6)$ kernel is the symmetrized sum of diagrams (with coefficients
that we have not determined) shown in Fig.~10. We also anticipate that
kernels for general colored channels can be obtained by determining the
appropriate color factor for each term relative to the color zero channel.

\subhead{6. SCALE DEPENDENCE}

{}From the construction it is, of course, clear that we will always obtain
kernels that are (transverse momentum) scale-invariant. It is possible that
these kernels have a fundamental relationship to the massless Regge region
S-matrix. They may also be conformally invariant\cite{lip} (this is being
studied). However, for physical applications we must input a scale. The
simplest possibility, which we are not interested in here, is that the gluon
becomes massive and this mass is added to all the transverse momentum
propagators. Our interest is to input the off-shell renormalisation scale of
QCD so that the evolution of the coupling, i.e. $g^2 /4\pi \to
\alpha_s(Q^2)$ somehow enters the formalism. This is non-trivial
since we can expect that all the possible transverse momentum scales in a
diagram will be involved in the scale-breaking.

Fadin and Lipatov have already calculated\cite{fad} the full trajectory
function (that is the disconnected piece) in the next-to-leading log
approximation in which we expect the $O(\alpha_s^2)$ kernel to appear. In
addition to the off-shell renormalization effects, they find that the
leading-log $ln s$ factors arise from $ln [s / k_{\perp}^2]$ where the
$k_{\perp}$ may be, essentially, any internal transverse momentum. As a
result the diagrams we have constructed contribute also with additional
internal logarithm factors, as illustrated in Fig.~11.The important feature
from our perspective is, however, that the diagram structure we have
anticipated is what is found. This encourages us to try to determine the
analagous logarithms that will occur as scale-breaking is introduced in the
remainder of the kernel. It appears that we may be able to do this by an
extension of the Ward identity plus infra-red finiteness analysis. As a
``preliminary'' result we note that the number of new diagrams do indeed
seem to match the number of new conditions.

If we define new logarithmic vertex functions as in Fig.~12 then, as
illustrated in Fig.~13, there are three kinematically distinct disconnected
diagrams. There are also, as illustrated in Fig.~14, nine further
kinematically distinct new diagrams. For this set of nine diagrams there are
four Ward identity and four infra-red finiteness constraints to determine
the relative weights. There are three forms of divergence generated by
further integration and so the relative weight of all the new disconnected
pieces should be determined by overall infra-red finiteness.

As a matter of principle it is, perhaps, not clear that the full kernel
should contain only (generalised) transverse momentum diagrams. Before we
can adequately discuss this we need to carry out the analysis we have
suggested. The component reggeon vertices already calculated by Fadin and
Lipatov\cite{fal}, that will go into their calculation of the full kernel,
are certainly complicated. Nevertheless the reggeon interaction that
provides the kernel is only a single partial-wave of the full 2-2 reggeon
scattering amplitude. It is this partial-wave projection which combines with
unitarity, as we have outlined, to determine that the infra-red behaviour of
the kernel must be described by transverse momentum diagrams. If this can
smoothly match with the asymptotic freedom of the coupling then our
diagrammatic description, including scale-breaking, should be sufficient.

\subhead{7. THE FUTURE}

We have already studied the $O(g^4)$ scale-invariant 2-2 kernel in some detail
and a paper describing various properties, including the leading eigenvalue,
will appear shortly. Our hopes for the future include

\begin{itemize}

\item{ Determination of the complete $O(\alpha_s^2)$ kernel, as we
have outlined.}

\item{To understand how the scaling violations produced by the
$\ln k^2_{\perp}$ factors combine with $\alpha_s(Q^2)$ to give simultaneous
evolution in $Q^2$.}

\item {Obtaining, perhaps, an all-orders ``Reggeon Field Theory'' describing
simultaneouous evolution in $ln {1 \over x}$ and $Q^2$.}

\item{ Input (massless) quarks to study\cite{arw2} the ``soft Pomeron'' and
confinement at small $ k_{\perp}$ .}

\end{itemize}

\vfill\eject

\vfill\eject

\centerline{\bf Figure Captions}

\begin{description}

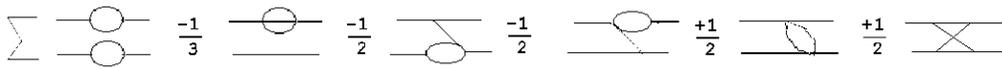
\item[Fig.~1] Diagrammatic representation of the $O(g^2)$ kernel.

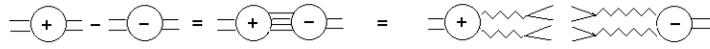
\item[Fig.~2] Regge pole contribution in the four-particle unitarity integral.

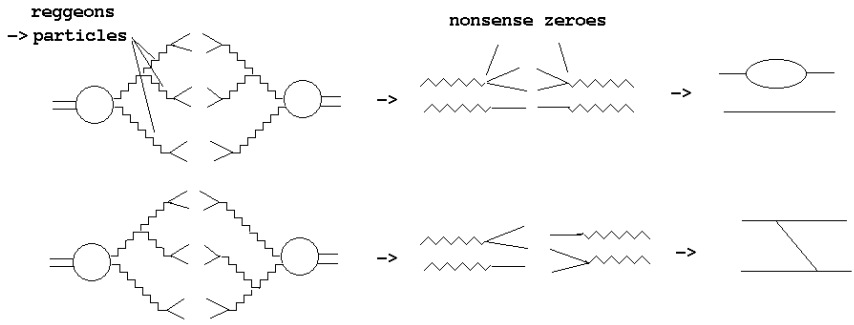
\item[Fig.~3] The generation of elements of the $O(g^2)$ kernel by three
reggeon nonsense states.

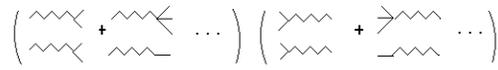
\item[Fig.~4] The combination of nonsense couplings to give the transverse
momentum diagrams generated by the four-particle nonsense state

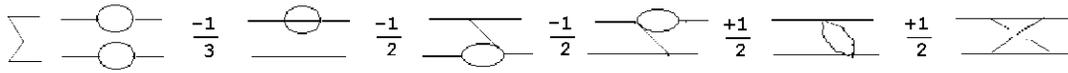
\item[Fig.~5] Diagrammatic representation of the four-particle $O(g^4)$ kernel

\item[Fig.~6] The contribution to the $O(g^4)$ kernel from iteration of the
two-particle nonsense state

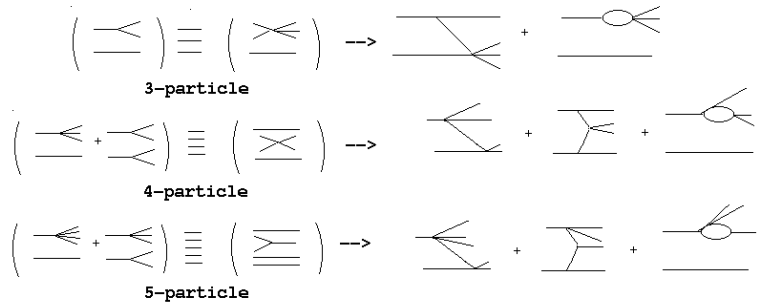
\item[Fig.~7] Nonsense states producing the $O(g^4)$ (2-4) kernel.

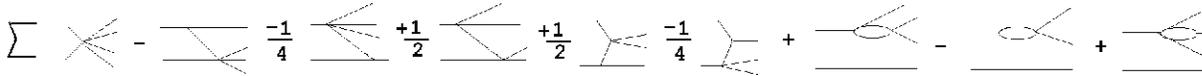
\item[Fig.~8] Diagrammatic representation of $K^{(4)}_{2,4}$.

\item[Fig.~9] The generation of transverse momentum diagrams for the
$O(g^{2N})$ kernel.

\item[Fig.~10] The $O(g^6)$ kernel generated by five-particle nonsense states.

\item[Fig.~11] Additional logarithms in the trajectory function diagrams.

\item[Fig.~12] New vertices - including logarithms.

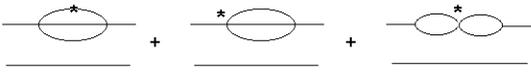
\item[Fig.~13] Disconnected (trajectory function) diagrams involving
logarithmic vertices.

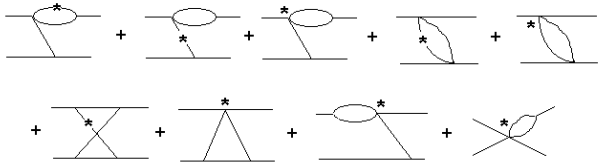
\item[Fig.~14] The nine additional diagrams involving logarithmic vertices.

\end{description}

\end{document}